\documentstyle[preprint,aps]{revtex}
\newcommand{\be}{\begin{equation}} \newcommand{\ee}{\end{equation}}
\newcommand{\bea}{\begin{eqnarray}}\newcommand{\eea}{\end{eqnarray}}

\begin{document}
\draft
\preprint{IP/BBSR/93-57}
\title { Degenerate Topological Vortex solutions from a generalized Abelian
Higgs Model with a Chern-Simons term }
\author{Pijush K. Ghosh\cite{mail}}
\address{Institute of Physics, Bhubaneswar-751005, INDIA.}
\maketitle
\begin{abstract}
 We consider a generalization of the abelian Higgs model with a Chern-Simons
term by modifying two terms of the usual Lagrangian. We multiply a
dielectric function with the Maxwell kinetic energy term and incorporate
nonminimal interaction by considering generalized covariant derivative.
We show that for a particular choice of the dielectric function
this model admits topological vortices satisfying Bogomol'nyi
bound for which the magnetic flux is not quantized
even though the energy is quantized. Furthermore, the vortex solution
in each topological sector is infinitely degenerate.
\end{abstract}
\narrowtext

\newpage
 In the last few years, the vortex solutions[1-10] in the
abelian Higgs model have received considerable attention
in the literature because of their possible relevance in the context of
cosmic strings as well as planar condensed matter systems. The general
feature of the topological vortices, both charged and neutral, in gauge
theories known to date is that the magnetic flux and the energy of the
vortices are both quantized. Furthermore, the vortex solution
in each topological sector is nondegenerate. Our purpose of this letter
is to show that a field theoretical model admitting topological
vortex solution can be constructed within the framework of
abelian Higgs model for which magnetic flux need not necessarily
be quantized even though the energy is quantized.
Further, this model admits infinitely degenerate vortex
solutions in each topological sector.

 We consider a generalization
of the abelian Higgs model with a Chern-Simons (${\bf CS}$) term, in
which we have a ``dielectric function" multiplying the
Maxwell term and an extra gauge invariant nonminimal contribution to
the covariant
derivtive. Specifically we are interested in the case where dielectric
function depends on the Higgs field. For a particular choice of the
dielectric function, this model admits both topological and nontopological
charged vortex
solutions obeying Bogomol'nyi bounds\cite{bogomol'nyi} for which the
magnetic flux, and hence the charge and the angular momentum need not
necessarily be quantized. The topological vortex solutions are infinitely
degenarate in each sector and these degenerate vortex
solutions in a particular sector differ from each other by flux,
charge and
the angular momentum.
In particular, the topological
vortex solutions in each sector are characterized by
energy $E={{\pi \kappa^2 n}
\over {2 e^2}}$, flux $\Phi={{2 \pi} \over e} (n-\beta)$,
angular momentum $J={{\pi \kappa}
\over e^2} (\beta^2-n^2)$ and charge
$Q= - \kappa \Phi$ ( where $n$ is the winding number,
$\kappa$ is the coefficient of the
{\bf CS} term, $e$ is a coupling constant and $\beta$ is a parameter
describing the solution). Using the sum rules\cite{khare277} for these
topological vortices we find that
$\beta$ is restricted as, ${1 \over 4} < \beta < n$.

 We first define our theory by writing the Lagrangian density
\be
{\cal L} \ = \ - \ {\ 1 \over 4} G({\mid \phi \mid})  F_{\mu \nu}  F^{\mu \nu}
+ \  {1 \over 2} D_{\mu} {\phi} (D^{\mu}{\phi})^* \
+ \ {\kappa \over 4}  \epsilon^{\mu \nu \lambda}  A_{\mu}  F_{\nu \lambda}
- \ V({\mid \phi \mid})
\label{eq:rl}
\ee
where $G({\mid \phi \mid})$ is the scalar field dependent
dielectric fuction and
the generalized covariant derivative is given by
\be
 D_\mu \phi \ = \ (\partial_{\mu}  \ - \ i e A_\mu
 - \ {{\ i g} \over 4} G(\mid \phi \mid) \epsilon_{\mu \nu \lambda}
F^{\nu \lambda})\phi
\label{eq:cd1}
\ee
\noindent Our notation is $F_{\mu \nu}={\partial_\mu}
{A_\nu} -{\partial_\nu} {A_\mu}$,
$\mu =(0,1,2)$
, $g_{\mu \nu}=diag (1,-1,-1)$, $c= \hbar =1$ and $\epsilon_{012}=1$.

 With a choice of symmetry breaking potential and using the
covariant derivative (\ref{eq:cd1}), both  mass term for the gauge field and
the ${\bf CS}$ term can be generated via spontaneous symmetry
breking({\bf SSB})
mechanism\cite{paul87}.
In fact the nonminimal part of the covariant derivative generates
the {\bf CS} term after {\bf SSB}. Also it is interesting to note that for
$G(\mid \phi \mid)$=1
the nonminimal part can be interpreted as anomoulous magnetic
moment\cite{kogan}. This is due to the fact that in 2+1 spac-time
Dirac matrices obey
$SO(2,1)$ algebra and Pauli coupling can be incorporated in the
generalized covariant derivative even for the scalar field without
mentioning the spin degrees of freedom.

 The modification to the Maxwell kinetic term can be can be viewed
as an effective action for a system in a medium described by a suitable
dielectric function. In fact, soliton bag models\cite{friedberg}
of quarks and gluons
are described by Lagrangian, where such a dielectric function is
multiplied with the Maxwell kinetic energy term. Also in certain
supersymmetric theories with a non-compact gauge group\cite{hull}, such a
nonminimal kinetic term was necessary in order to have a sensible gauge
theory. In context of vortex solution, this non-minimal coupling
is interesting because of the existence of Bogomol'nyi bounds for
a more general form of the scalar potential\cite{nam,lohe}.
Lee et. al.\cite{nam} considered
the lagrangian (\ref{eq:rl}) with ${\kappa=0}$ and without the nonminimal
contribution to the covariant derivative (\ref{eq:cd1}) and have shown
that the model admits topological as well as nontopological static
self-dual neutral vortex solutions. Recently, Torres\cite{torres}
considered
the lagrangian (\ref{eq:rl}) with ${\ G(\mid \phi \mid)}=1$ and obtained static
minimum energy nontopological vortex configuration
for a simple ${\mid \phi \mid}^2$ potential. As a natural extension, we
consider the effect of both the dielectric function and the generalized
covariant derivative for arbitrary ${\ G(\mid \phi \mid)}$
and study the Bogomol'nyi
limit for topological vortex solution.

 The equations of motion for the Lagrangian in equation (\ref{eq:rl}) are
\be
D_{\mu}D^{\mu}\phi \ = \ - \ 2 {{\partial V(\mid \phi \mid)} \over
{\partial \phi^*}} \ - {\ 1 \over 2} {{\partial G(\mid \phi \mid)} \over
{\partial \phi^*}} F_{\mu \nu} F^{\mu \nu}
 \ - \ {\ g \over 2 e} {{\partial G(\mid \phi \mid)} \over {\partial \phi^*}}
\epsilon^{\mu \nu \lambda} J_\mu F_{\nu \lambda}
\label{eq:sfr}
\ee
\be
\epsilon_{\mu \nu \lambda} \ {\partial^\mu} \ [G(\mid \phi \mid) \ (F^\lambda
\ + {\ g \over 2 e} J^\lambda)] \ = \ J_\nu \ - \ \kappa F_\nu
\label{eq:gfr}
\ee
where the dual field
$F_\mu \ = \ {\ 1 \over 2} \epsilon_{\mu \nu \alpha} \  F^{\nu \alpha}$
and the conserved current $J_\mu$ is
\be
J_\mu \ = \ - \ {{\ i e} \over 2} [\phi^* D_\mu \phi \
- \ \phi (D_\mu \phi)^*]
\label{eq:cdr}
\ee

 The energy momentum tensor $T_{\mu \nu}$ is obtained by varying the curved
space form of the action with respect to the metric
\bea
T_{\mu \nu} \ &  = & \ G({\mid \phi \mid}) \ [1-
{\ g^2 \over 4} G({\mid \phi \mid}) {\mid \phi \mid}^2] \ [F_\mu F_\nu-
{\ 1 \over 2} g_{\mu \nu} F_\alpha F^\alpha]\nonumber \\
& & + { \ 1 \over 2} [\bigtriangledown_\mu \phi
(\bigtriangledown_\nu \phi)^*+{\bigtriangledown_\nu \phi}
({\bigtriangledown_\mu \phi})^*]\nonumber \\
& & - \ g_{\mu \nu} [{\ 1 \over 2} {\mid {\bigtriangledown_\alpha} \phi
\mid }^2 -
V ({\mid \phi \mid}) ]
\label{eq:emtr}
\eea
\noindent where $\bigtriangledown_\alpha \ = \ \partial_\alpha \ - \
i e A_\alpha$ includes
only the gauge potential contribution.

 Now notice that the first order equation
\be
J_\nu \ = \ k F_\nu
\label{eq:fgf}
\ee
solves the gauge field equation (\ref{eq:gfr}) for
arbitrary $G({\mid \phi \mid})$
provided the following relation among the coupling constant holds
\be
g \ = \ - \ {\ 2e \over \kappa}
\label{eq:cc}
\ee
\noindent Like pure {\bf CS} higgs theory\cite{hong}, the zero
component of (\ref{eq:fgf}), i. e, Gauss law implies that the
solution with charge $Q$ also carries
magnetic flux ${\Phi} \ = \ - \ {Q \over \kappa}$. It should be noted that for
${G({\mid \phi \mid}) \neq 0}$ equation (\ref{eq:fgf}) is
essentially different from
that of corresponding equation for pure ${\bf CS}$ vortices as $D_\mu$
receives contribution from the nonminimal part also. In particular,
using equations (\ref{eq:cdr}) and (\ref{eq:cc}) the gauge field
equation (\ref{eq:fgf}) can be
rewriten as
\be
\kappa \ F_\mu =
J_\mu \ = \ (1 - {e^2 \over \kappa^2} G(\mid \phi \mid)
{\mid \phi \mid}^2)^{-1} \tilde{J_\mu}
\label{eq:nfgf}
\ee
where $\tilde{J_\mu}$ receives only minimal contribution,
\be
\tilde{J_\mu} \ = \ - \ {{\ i e} \over 2} [\phi^* \bigtriangledown_\mu \phi \
- \ \phi (\bigtriangledown_\mu \phi)^*]
\label{eq:ncdr}
\ee

 Now we seek vortex solutions in the system described by
the equations (\ref{eq:sfr}), (\ref{eq:emtr}) and (\ref{eq:nfgf}),
when the relation (\ref{eq:cc}) is satisfied.
We choose the ansatz for rotationally symmetric solution of winding
number $n$
\be
\vec{A(r)} \ = \ - \ {\hat \theta} {{a(r) \ - \ n} \over {e r}} ,
A_0 (r) \ = \ {\ k \over e} h(r) ,
\phi(r) \ = \ {\ k \over e} f(r) e^{- i n \theta}
\label{eq:ansatz}
\ee
\noindent After substituting the ansatz (\ref{eq:ansatz}),
equations (\ref{eq:sfr}) and (\ref{eq:nfgf}) can be reduced to
\be
{\ 1 \over r} \ [1 \ - \ G(f) f^2] a^\prime \ + \ k^2 f^2 h
\ = \ 0
\label{eq:rs1}
\ee
\be
r [1 \ - \ G(f) f^2 ] h^\prime \ + \ a f^2 \ = \ 0
\label{eq:rs2}
\ee
\be
{\ 1 \over r} {{\partial} \over {\partial r}} (r {{\partial f} \over
{\partial r}}) \  +  {\ k^2 f \over {\ (1\ - \ G(f) f^2)^2}} (1 \
+ \ {\ f^3 \over 2} {{\partial G(f)} \over {\partial f}}) (h^2 \ -
{\ a^2 \over {\ k^2 r^2}})  =  \ {\ e^2 \over k^2} {{\partial V}
\over {\partial f}}
\label{eq:rs3}
\ee
\noindent where prime denotes differentition with respect to r.
The energy functional that is obtained
from equation (\ref{eq:emtr}) for the ansatz (\ref{eq:ansatz}) is
\bea
E \ & = & \ {\ k^2 \over 2 e^2} {\cal \int} d^2 x \{ \ G(f) (1 \ - G(f) f^2)
[(h^\prime)^2 \ + \
({\ a^\prime \over k r})^2] \ + \ (k h f)^2\nonumber \\
& & + \ (f^\prime)^2 \ + \
({f a \over r})^2 \ + \
{\ 2 e^2 \over k^2} V(f) \ \}
\label{eq:rse}
\eea

 We choose $G(f)$ to be
\be
G(f)=f^{\ -2} - C_{0} \ f^{- 2} \ (1-f^2)^{\ 1- \gamma}
\label{eq:df}
\ee
\noindent where $\gamma$ is a real number and $C_{0}$ is a
positive constant. For this choice of $G(f)$, the energy functional
can be rearranged using Bogomol'nyi trick
\bea
E \ &  = & \ {\ k^2 \over {\ 2 e^2}} {\cal \int} d^2 x \{ [f^\prime
\ \pm \ {{\ f a} \over {\ {C_{0}}^{1 \over 2} r}} (1-
f^2)^{{\gamma \ - \ 1} \over
2}]^2 \nonumber \\
& & + C_{0} f^{\ -2} (1-f^2)^{\ 1- \gamma} \ [{{\ a^\prime} \over
{\kappa  r}} \ \mp \
{{\kappa} \over {\ C_{ 0}^{\ 3 \over 2} (1+\gamma)}} f^2 (1-
 f^2)^{{\ 3 \gamma - 1} \over 2}]^2 \nonumber \\
& & +{\ 2 e^2 \over k^2} V(f)-
{{\kappa^2} \over {\ C_{ 0}^2 (1+\gamma)^2}}
f^2 (1-f^2)^{\ 2 \gamma}\} \ \pm \
{{2 \pi \kappa^2} \over {\ C_{ 0}^{\ 1
\over 2} (1+\gamma) e^2}} \ [ R(\infty) \ - \ R(0) ]\nonumber \\
R(r) \ & = & \ a (1-f^2)^{{1+\gamma} \over 2}
\label{eq:be}
\eea
\noindent where $h(r)$ and $h^\prime(r)$ have been eliminated from
equation (\ref{eq:rse})
using equations
(\ref{eq:rs1}) and (\ref{eq:rs2}). When $\gamma$ is an odd integer
then there
is a lower bound on the
energy provided we choose the scalar potential
\be
V(f)={{\kappa^4} \over {\ 2 C_{0}^2 e^2 (1+\gamma)^2}} f^2 (1
-f^2)^{\ 2 \gamma}
\label{eq:sp}
\ee
For arbitrary $\gamma$ and above choice of $V(f)$, also lower bound on
the energy exists with restriction on f(r), $0 \leq f \leq 1$. The lower
bound on energy is saturated when the following Bogomol'nyi equations
are satisfied
\be
f^\prime = \ \pm
\  \ {{\ f a} \over {\ {C_{0}}^{1 \over 2} r}} (1-
f^2)^{{\gamma - 1} \over
2}
\label{eq:fbe1}
\ee
\be
{{\ a^\prime} \over
{\kappa  r}} \ = \ \mp  \
{{\kappa} \over {\ C_{0}^{\ 3 \over 2} (1+\gamma)}} f^2 (1-
 f^2)^{{\ 3 \gamma - 1} \over 2}
\label{eq:fbe2}
\ee
\noindent One can easily check that these two first order differential
equations are consistent with the second order differential equation
(\ref{eq:rs3}). At the Bogomol'nyi limit, two diagonal elements of
the energy momentum
tensor other than $T_{\ 0 0}$, i.e, $T_{\ r r}$ and $T_{\theta \theta}$
vanishes. The off-diagonal elment $T_{\ t \theta}$ is nonvanishing,
$T_{\ t \theta}=- {\kappa \over {\ e^2 r^2}} a^\prime a$,
implying that the solution to the Bogomol'nyi equations carry finite
angular momentum
$J={{\pi \kappa} \over e^2} \ [a(\infty)^2-a(0)^2)]$, for well behaved $a(r)$.
Note that the following scale transformation
\be
a \rightarrow C_{0}^{\ 1 \over 2} \ a \  , \
r \rightarrow C_{0} \ r \ , \ f \rightarrow \ f
\label{eq:st}
\ee
\noindent eliminates $ C_{ 0}$ from the equations (\ref{eq:fbe1})
and (\ref{eq:fbe2}). The
decoupled second order equation for the $f(r)$ is
\be
f^{\prime \prime}+{\ f^\prime \over r}+
{\kappa^2 \over {\ \gamma+1}} f^3 (1-f^2)^{\ 2 \gamma -1}-
{{\ f^\prime}^2 \over f}+({\gamma-1}) {\ {{f^\prime}^2 f} \over {\ 1-f^2}}=0
\label{eq:ds}
\ee
\noindent where the scale transformation (\ref{eq:st}) have been performed.
Equation (\ref{eq:ds}) is highly nonlinear and we do not have
an analytical solution for
it. However we can obtain asymptotic solution for this equation. We
do so while discusing vortex solution for different choice of $\gamma$.

 I. \ \ $\gamma = 0 \ $, i.e, $G=(1-C_{0}+C_{0} f^2) f^{\ -2}$ \ :
For $C_{0 }=1$, i.e, $G=1$; this model reduces to to that of
considered by Torres\cite{torres}. For arbitrary $C_{0} (C_{0} > 0)$
Bogomol'nyi equations of our
model can be mapped into corresponding equations of
Torres model upto a scale transformation of the variables. For this case
only nontopological vortices exists.

 II. \ \ $\gamma=1$, i.e, $G=(1-C_{0}) f^{\ -2}$ :
For $C_{0}=1$, i.e, $G=0$, this model
reduces to pure ${\bf CS}$ Higgs theory\cite{hong}. Equations (\ref{eq:fbe1}),
(\ref{eq:fbe2}), (\ref{eq:sp})
reproduces the Bogomol'nyi equations and the potential of pure ${\bf CS}$
Higgs theory respectively. For arbitray $C_{0} \ (C_{0} \ > \ 0)$
also, these Bogomol'nyi equations are exactly same
as that of corresponding
Bogomol'nyi equations for pure ${\bf CS}$ theory provided the
scale transformation (\ref{eq:st}) is performed. This implies that
the Bogomol'nyi equations of our model
can be mapped into the
Bogomol'nyi equations of pure ${\bf CS}$ higgs theory upto a scale
transformation of the variables.

 III. \ \ $\gamma = 3 \ $: \ \
\noindent The finiteness of the energy can be ensured by requireing
either (i)$a(\infty)= \ - \alpha$, $f(\infty)=0$ or
(ii)$a(\infty)={\beta}$, $f(\infty)=1$. Here $\alpha$
and ${\beta}$ are two real positive constants.
Further, on demanding nonsingular field variables, boundary condition at
the origin gets fixed as (iii)$ a(0)=n \ , \  f(0)=0$.
The boundary condition (i) corresponds to nontopological vortex solution
while the boundary condition (ii) corresponds to topological vortex
solution. For simplicity we will discuss in detail only topological
vortex solutions(without loss of generality we choose $C_{0}=1$); the
nontopological as well as nonrelativistic vortex solutions and
other related aspects of this model will be discussed elsewhere\cite{pijush}.
Since the solutions for $n$ and $-n$ are related by the transformation
$f \rightarrow f$, $a \rightarrow -a$; we consider only the case
$n > 0$. The behaviour at small distances is given by
\be
f(r) = \ B (\kappa r)^n \ - \ { 1 \over 2} B^3 (\kappa r)^{ 3 n} \ + \
{\bf O}( \ (\kappa r)^{ 3 n+2} \ )
\label{eq:ai3}
\ee
\bea
a(r)  = n  - \ {\ B^2 \over {8 (n+1)}} (\kappa r)^{\ 2 n+2}
 \ + \ { {5 B^4 (\kappa r)^{ 4 n+2}}  \over {\ 8 (2
n \ + \ 1) }}\ + \ {\bf O}( \ (\kappa r)^{6 n+2})
\label{eq:ai4}
\eea
\noindent We obtain large distance behaviour as following
\be
f(r) \ = \ 1 + {D \over (\kappa r)^{2 {\beta}}} +
{{3 D^2} \over {2 (\kappa r)^{4 {\beta}}}} \ +
\ {\bf O} ( \ ({1 \over {\kappa r}})^{6 \beta})
\label{eq:ai7}
\ee
\be
a(r) \ = \ {\beta} + {{2 D^4} \over {(4 {\beta}-
1) (\kappa r)^{8 {\beta}-2}}} + {{16 D^5 } \over {(5 {\beta} \ - \
1) (\kappa r)^{10 {\beta}-2}}} \
+ \ {\bf O} ( \ ({1 \over \kappa r})^{12 \beta \ - \ 2})
\label{eq:ai8}
\ee
\noindent Note that the large distance behaviour of the scalar field and the
gauge field for these topological vortices are of
semi-local\cite{semi-local} type, i.e, they fall off obeying
power law.

 It is remarkable to note that when
the scalar field $f(r)$ attains its asymmetric vaccum value at large
distances, $a(r)$ does not vanish; a feature not known for topological
vortices. The novel consequnce is that even for topological
vortices the magnetic flux, and hence the charge and the angular momentum,
need not necessarily be quantized; while the energy, as evident
from (\ref{eq:be}), is quantized. In particular, the topological
vortex solutions are characterized by energy $E={{\pi \kappa^2 n}
\over {2 e^2}}$, flux $\Phi={{2 \pi} \over e} (n-\beta)$, charge
$Q= - \kappa \Phi$ and angular momentum $J={{\pi \kappa}
\over e^2} (\beta^2-n^2)$. Since for each $n$, there is a set
of solutions paramterized by $\beta$; the solutions are degenerate
in each topological sector
and each solution differ from each other by charge, flux and angular
momentum. At this point its worthwhile to ask whether $\beta$ can
take any positive value or bounded from both up and below. To our
surprise the latter is indeed true and infact ${1 \over 4} <\beta
< n$. The lower bound on $\beta$ is due to the fact that
the second term in (\ref{eq:ai8}) is subleading compare to the first
term only when $\beta > {1 \over 4}$. The upper bound follws from
the sum rules for topological vortices first
obtained by Khare\cite{khare277}. Using equations (\ref{eq:fbe1}),
(\ref{eq:fbe2}) and the boundary conditions (ii) and (iii),
we find that the first two sum rules are
\be
n - \beta = {1 \over 4} \int_{0}^{\infty} r dr f^2 ( 1 - f^2 )^4
\label{eq:sr4}
\ee
\be
n^2 - \beta^2 = {1 \over 8} \int_{0}^{\infty} r dr ( 1 - f^2 )^4
\label{eq:sr5}
\ee
\noindent For both the sum rules (\ref{eq:sr4}) and (\ref{eq:sr5})
the right hand side is positive definite and this imply that $\beta \leq n$.
When $\beta=n$ then the trivial solution $f(r)=1$, $a(r)=n$ with
zero energy, charge, flux and angular momentum satisfies the equations
(\ref{eq:fbe1}) and (\ref{eq:fbe2}). So the upper bound is exact.
Since $\beta$ is allowed to take any value between ${1 \over 4}$ and $n$,
the solutions in each topological sector are infinitely degenarate. To
show that topological vortex solution exist for different $\beta$, we
have integrated (\ref{eq:fbe2}) and (\ref{eq:ds}) numerically
and  plotted $f(r)$ (solid line)
and $a(r)$ (dashed line) in Fig. 1 for $n=1,2$ with different values
of $\beta$.

Following comments are in order. For $\gamma \geq 3$, the Bogomol'nyi
bounds for the nontopological vortex solutions can be obtained in this model
\cite{pijush}.
Using sum the rule technique, lower bound on the magnetic flux can be put
which are in agreement with numerical calculation\cite{pijush}. Furthermore,
in the nonrelativistic limit of this model, self-dual soliton
solution exists saturating the lower bound at some finite value of
the energy unlike
pure ${\bf CS}$\cite{pi} and
Maxwell {\bf CS}\cite{dunne} theory. For the
whole class of the dielectric function for which nontopological vortex
solution exist in the relativistic theory, the charge density solves
the same Liouville equation in the nonrelativistic limit\cite{pijush},
which is completely integrable. All these issues will be discussed in
detail in a forthcoming article\cite{pijush}

 The Bogomol'nyi equations we obtain are quiet
different in nature from the corresponding equations for vortices in
other well known models. Naturally it is worthwhile to study whether the usual
technique for showing the uniqueness and existence of soliton
solution goes through in this case or not. Also it would be interesting
to know the total number of independent zero modes present in this model.
Above all, the most interesting thing it would be if
this model can be realized in any planar condesed matter system where
dielectric function plays major role.

\acknowledgements
 I thank Prof. Avinash Khare for valuable discussions and critically
going through the manuscript. I thank Munshi Golam Mustafa for helping
me in numerical computation.

\begin{figure}
\caption{ A plot of $f(r)$ (solid line) and $a(r)$ (dashed line)
for (I) $n=1, \beta=0.78$; (II) $n=2, \beta=1.91$; (III) $n=2, \beta=1.53$
and (IV) $n=2, \beta=1.23$}
\end{figure}

\end{document}